\documentclass[prb,twocolumn,showpacs,preprintnumbers]{revtex4-1}
\usepackage{graphicx}
\usepackage{epstopdf}
\usepackage{array,multirow,amsmath,amsfonts}
\usepackage{color}

\newcolumntype{Y}{>{\centering\arraybackslash}X}
\newcolumntype{C}[1]{>{\centering\arraybackslash}p{#1}}

\begin{document}
\preprint{}

\title{
Dimensionality-strain phase diagram of strontium iridates
 }

\author{Bongjae Kim$^{1}$}
\author{Peitao Liu$^{1,2}$}
\author{Cesare Franchini$^{1}$}

\affiliation{$^{1}$University of Vienna, Faculty of Physics and Center for Computational Materials Science, Vienna, Austria}
\affiliation{$^{2}$Shenyang National Laboratory for Materials Science, Institute of Metal Research, Chinese Academy of Sciences, Shenyang 110016, China}

\date[Dated: ]{\today}

%%%%%%%%%%%%%%%%%%%%%%%%%%%%%%%%%%%%%%%%%%%%%%%%%%%%%%%%%%
\begin{abstract}
The competition between spin-orbit coupling, bandwidth ($W$) and electron-electron interaction ($U$) makes iridates highly susceptible to small external perturbations, which can trigger the onset of novel types of electronic and magnetic states. Here we employ {\em first principles} calculations based on density functional theory and on the constrained random phase approximation to study how dimensionality and strain affect the strength of $U$ and $W$ in (SrIrO$_3$)$_m$/(SrTiO$_3$) superlattices. The result is a phase diagram explaining two different types of controllable magnetic and electronic transitions, spin-flop and insulator-to-metal, connected with the disruption of the $J_{eff}=1/2$ state which cannnot be understood within a simplified local picture.
\end{abstract}
%%%%%%%%%%%%%%%%%%%%%%%%%%%%%%%%%%%%%%%%%%%%%%%%%%%%%%%%%


\maketitle

 \section{Introduction}

 Many interesting phenomena found in transition metal oxides are explained by the competition of intertwined energy scales usually parameterized as electronic correlation ($U$), bandwidth ($W$), spin-orbit-coupling (SOC) and crystal field splitting. In 3$d$ oxides, $U$ typically acts as a leading parameter and this sets the ground for a variety of interesting effects such as Mott-like insulator-to-metal transition (IMT), superconductivity, and spin/orbital/charge orderings~\cite{Imada1998}.
In the heavier 4$d$ and 5$d$ transition metal oxides the Mott paradigm is largely attenuated owing to the stronger SOC and the broader spatial extension of the $d$ orbitals (larger $W$):  $U$ is not the dominant factor anymore and the competing balance between similar energy scales ($U$, $W$, and SOC) promotes the onset of a novel and often exotic physics~\cite{Watanabe2010,Martins2011,Watanabe2013,Witczak-Krempa2014}. Strontium-based iridates represent the archetypal playground for these uncommon behaviors. The most notable example is the relativistic-Mott insulating $J_{eff}$=1/2 state associated with a canted antiferromagnetic planar ordering realized in Sr$_2$IrO$_4$~\cite{Cao1998,BJKim2008,BJKim2009}.

 %-----------------------------------------------------------------------
\begin{figure}[b]
\begin{center}
\includegraphics[width=85mm]{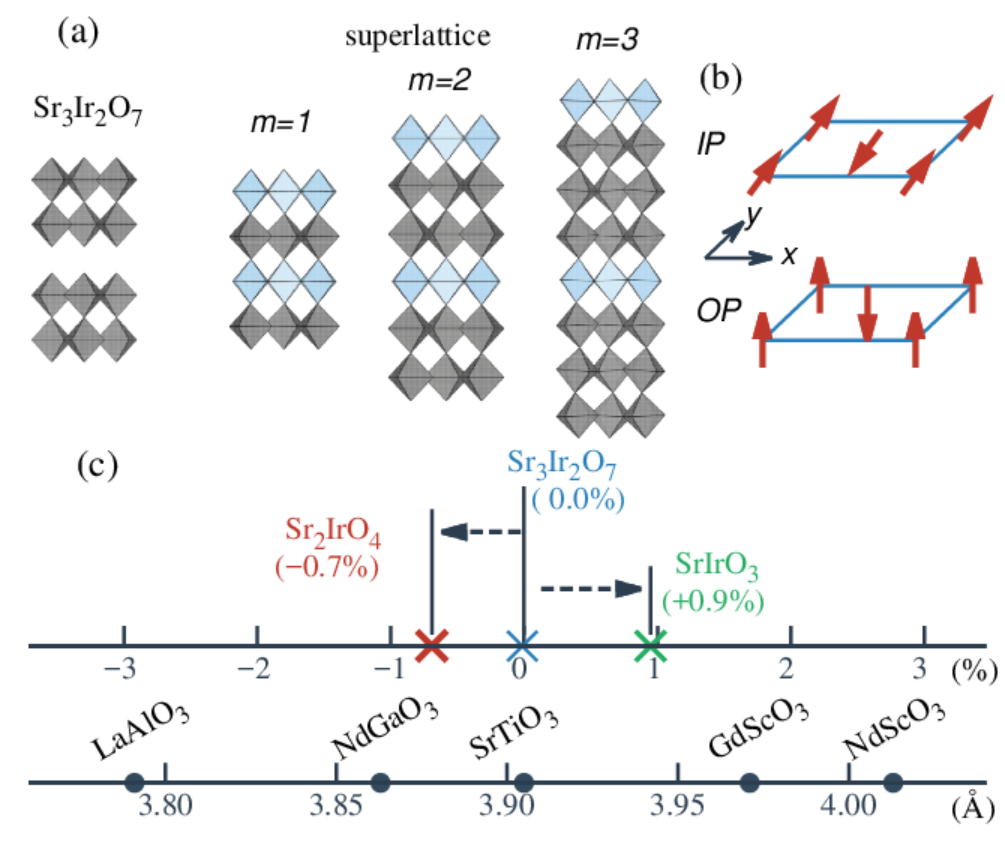}
\caption{ (Color online)
(a) Crystal structures of bilayer (Sr$_3$Ir$_2$O$_7$, $m=2$) and of (SrIrO$_3$)$_m$/(SrTiO$_3$) SL for different $m$.
IrO$_6$ and TiO$_6$ octahedra are depicted in gray and light gray (blue), respectively.
O and Sr atoms are not shown.
(b) Schematic description of the IP and OP magnetic orderings for one single IrO$_2$ layer.
(c) 2D lattice parameters of typical substrates and corresponding lattice mismatch between SrTiO$_3$
and RP strontium iridates
}
\label{fig1}
\end{center}
\end{figure}

 As the strength of the different physical interactions at play in the 5$d$ oxides is similarly small (about 1~eV), the electronic and magnetic ground state in these systems is expected to be highly susceptible to small changes of the interaction parameters ~\cite{Jackeli2009,Zeb2012,Matsuno2015,Witczak-Krempa2014,Sato2015,PLiu2015,BKim2016,BKim2017}.
 A realistic tuning of these interactions can be achieved using external perturbations such as doping and strain or via a change of the structural
 stacking~\cite{YKKim2014,YKKim2015,Hampel2015,delaTorre2015,PLiu2016,Cao2016,Moon2008,Yamasaki2016,Nichols2013,JLiu2013,Lupascu2014,JLiu2016,Ding2016}. For iridates, apart from doping~\cite{PLiu2016,delaTorre2015,Cao2016}, two approaches have been used often to tune the ground state properties: (i) variation of the degree of dimensionality through a modification of the structural stacking intrinsic to the Ruddlesden-Popper (RP) series~\cite{Beznosikov2000} Sr$_{m+1}$Ir$_{m}$O$_{3m+1}$ ($m=1,2,\cdots \infty$)~\cite{Moon2008,Carter2013,Zhang2013,Yamasaki2016}, and (ii) strain engineering~\cite{Nichols2013,JLiu2013,Lupascu2014,JLiu2016,BKim2016,BKim2017}.
 In the RP series, the degree of dimensionality is controlled by the number of IrO$_2$ layers ($m$) interleaved in the layered perovskite structure (see Fig.~\ref{fig1}  (a) for $m=2$ case). As $m$ increases, the system undergoes a dimensional crossover from the 2-dimensional (2D) limit ($m=1$, Sr$_2$IrO$_4$), characterized by the aforementioned magnetically canted relativistic-Mott state,
 to the 3-dimensional (3D) limit ($m=\infty$, SrIrO$_3$) which is nonmagnetic, (semi)metallic and exhibits nontrivial topological features~\cite{Carter2012,Zeb2012,JLiu2016}. The accepted picture is that with increasing $m$, also $W$ increases and this leads to a progressive disruption of the $J_{eff}$=1/2 phase~\cite{Moon2008,Zhang2013}. The first magnetic transition is observed already at $m=2$, manifested by a spin-flop transition from an in-plane (IP) to out-of-plane (OP) spin ordering caused by interlayer exchange coupling~\cite{JWKim2012} (Fig.~\ref{fig1} (b)). On the other hand, strain engineering uses the lattice mismatch between the strontium iridates and a given substrate,
 resulting in compressive and tensile strain usually up to 5\%  depending on the type of substrate material (Fig.~\ref{fig1} (c)).
 Optical studies indicate that strain effects induce changes in $U$, $W$ and SOC~\cite{Nichols2013,BKim2016}, as well as modifications of the strength of the magnetic
 interactions between Ir sites~\cite{Lupascu2014,PLiu2015,BKim2017}.
 However, a detailed explanation of these strain-induced changes and their repercussion on the $J_{eff}$=1/2 state remains still elusive.

%-----------------------------------------------------------------------

 The possibility to control in the same experiment both dimensionality and strain represents a viable route to enhance the spectrum of tuning opportunities. This has become feasible using superlattice (SL) structures thanks to the development of improved growing technique and the diverse choice of available substrate materials~\cite{Schlom2007}. A first example in this direction is the research of Matsuno {\it et. al.}  in which the authors have modulated the number of SrIrO$_3$ layers within SLs of (SrIrO$_3$)$_m$/(SrTiO$_3$) and demonstrated that it is possible to guide a tailored IMT from $m=1$ to $m=4$ by means of only structural assembly~\cite{Matsuno2015}.

 In this work, we aim to perform a computational experiment not yet realized in practice which consists in considering simultaneously the effect of both dimensionality and substrate strain in (SrIrO$_3$)$_m$/(SrTiO$_3$) SLs.  We do this fully {\it ab initio} by combining density functional theory (DFT) and the constraint random phase approximation (cRPA) calculations. The cRPA approach is employed to quantify the variation of $U$ as a function of $m$ and strain from a fully {\it ab initio} perspective. By means of DFT+SOC+$U$, we scrutinize the detailed structural changes as a function of $m$ and strain ($\leq$ 5\%) and their impact on $W$ and on the moment ordering.
 The result is a dimensionality-strain quantum phase diagram of strontium iridates constructed from first principles with {\it realistic} values of strain, $U$ and $W$. We show that $U$ and $W$ are largely dependent on dimension and strain and their accurate estimation is crucial to achieve a comprehensive interpretation of the electronic (IMT) and magnetic (spin-flop) transitions driven by $m$ and strain. Moreover, the possibility to selectively control the leading interactions in iridates allow for a direct assessment of the robustness and validity of the $J_{eff}$=1/2 local picture.

 \section{Calculational Details}

%====================================================================
\begin{table}[b]
\centering
\caption[]{
Calculated on-site Coulomb ($U_{ij}$) and exchange ($J_{ij}$) interaction values (in eV) within Ir-$t_{2g}$ orbitals based on the cRPA calculations for $m=1$ SL without any strain.
}
%\begin{tabular}{c|ccc|p{2.8cm}|ccc}
\begin{tabular}{C{0.9cm}|C{0.9cm}C{0.9cm}C{0.9cm}|C{0.9cm}|C{0.9cm}C{0.9cm}C{0.9cm}}
\hline\hline
  $U_{ij}$         & $xy$  & $xz$ & $yz$ &     $J_{ij}$    & $xy$  & $xz$ & $yz$ \\
\hline
  $xy$        & 2.43  & 1.79 & 1.63 &     $xy$   & -     & 0.25 & 0.24 \\
  $xz$        & 1.79  & 2.43 & 1.63 &     $xz$   & 0.25  & -    & 0.24 \\
  $yz$        & 1.63  & 1.63 & 2.05 &     $yz$   & 0.24  & 0.24 & -    \\
\hline
\end{tabular}
\label{crpa1}
\end{table}
%====================================================================

%====================================================================
\begin{table}[b]
\centering
\caption[]{
Calculated on-site Coulomb ($U_{ij}$) and exchange ($J_{ij}$) interaction values (in eV) within Ir-$t_{2g}$ orbitals based on the cRPA calculations for $m=2$ SL without any strain.
}
%\begin{tabular}{c|ccc|p{2.8cm}|ccc}
\begin{tabular}{C{0.9cm}|C{0.9cm}C{0.9cm}C{0.9cm}|C{0.9cm}|C{0.9cm}C{0.9cm}C{0.9cm}}
\hline\hline
  $U_{ij}$         & $xy$  & $xz$ & $yz$ &     $J_{ij}$    & $xy$  & $xz$ & $yz$ \\
\hline
  $xy$        & 2.09  & 1.49 & 1.42 &     $xy$   & -     & 0.22 & 0.21 \\
  $xz$        & 1.49  & 2.09 & 1.42 &     $xz$   & 0.22  & -    & 0.21 \\
  $yz$        & 1.42  & 1.42 & 1.83 &     $yz$   & 0.21  & 0.21 & -    \\
\hline
\end{tabular}
\label{crpa2}
\end{table}
%====================================================================

%====================================================================
\begin{table}[b]
\centering
\caption[]{
Calculated unscreened (bare) on-site Coulomb ($U^{bare}$) and exchange ($J^{bare}$) interaction values (in eV) for 0\% strain case with different $m$.
}
%\begin{tabular}{c|ccc|p{2.8cm}|ccc}
\begin{tabular}{C{1.8cm}|C{1.8cm}C{1.8cm}C{1.8cm}}
\hline\hline
  $             $         & $m=1$  &  $m=2$ & $m=3$ \\
\hline
  $U^{bare}$         & 8.70   &  8.70  & 8.51  \\
  $J^{bare}$         & 0.25   &  0.25  & 0.24 \\
\hline
\end{tabular}
\label{crpa3}
\end{table}
%====================================================================

  We performed {\it ab initio} electronic structure calculations using the projector augmented wave method employing the Vienna {\it ab initio} simulation package (VASP) \cite{Kresse1993,Kresse1996}. We adopted the generalized gradient approximation by Perdew-Burke-Ernzerhof (PBE) with a full treatment of relativistic effects (SOC) and including an on-site Hubbard $U$. We have used a plane-wave cutoff of 400 eV \cite{Perdew1996,Dudarev1998}. Benchmark calculations with a cutoff 600 eV do not lead to significant differences.
  Regarding the structure of the (SrIrO$_3$)$_m$/(SrTiO$_3$) superlattice, we took as reference in-plane lattice parameter the one of SrTiO$_3$,
  and varied it up to 4\% considering both tensile and compressive strain. Full atomic relaxation was performed for all studied $m$ at the corresponding
  IP lattice parameter using $U$=2~eV. We found that the structure is only marginally sensible on the specific value of $U$: by varying the $U$ from 0 to 2~eV, the structural changes are in the order of $10^{-3}$~\AA. Monkhorst-Pack $k$-meshes of 6$\times$6$\times$4, 6$\times$6$\times$2 and 6$\times$6$\times$3 are used for $m=1$, $m=2$, and $m=3$ superlattices, respectively. We note that for the $m=2$ system we have employed a twice larger supercell in order to allow for $a^-a^-c^+$ type tilting of the octahedra network along the $c$-direction.

%-----------------------------------------------------------------------
\begin{figure}[t]
\begin{center}
\includegraphics[width=85mm]{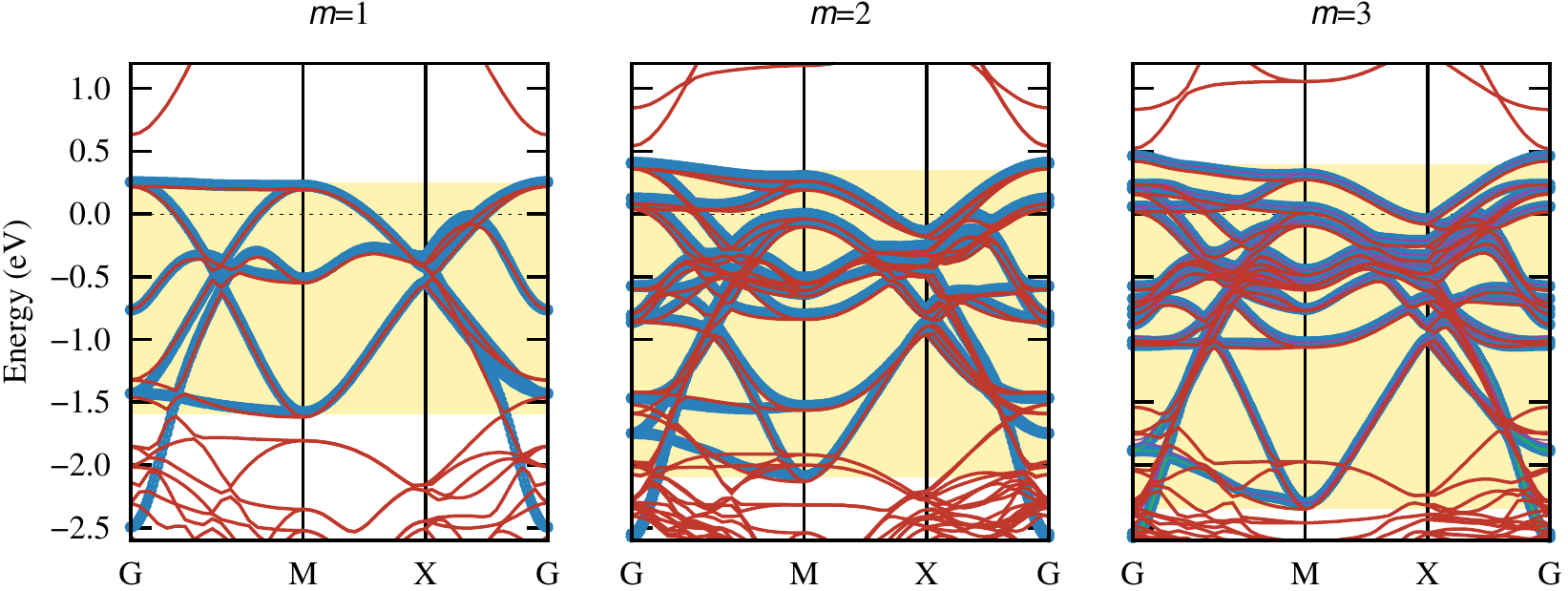}
\caption{(Color online)
Thin (red) lines denote PBE band structures of SLs with nonmagnetic setup, for different values of $m$.
Wannier-interpolated bands are shown with with thick (blue) lines. The shaded background (yellow) marks to the $t_{2g}$ bandwidths of the systems.
Substrate strain is set to 0\%.
}
\label{fig2}
\end{center}
\end{figure}
%-----------------------------------------------------------------------
%-----------------------------------------------------------------------
\begin{figure*}[t]
\begin{center}
\includegraphics[width=0.9\textwidth]{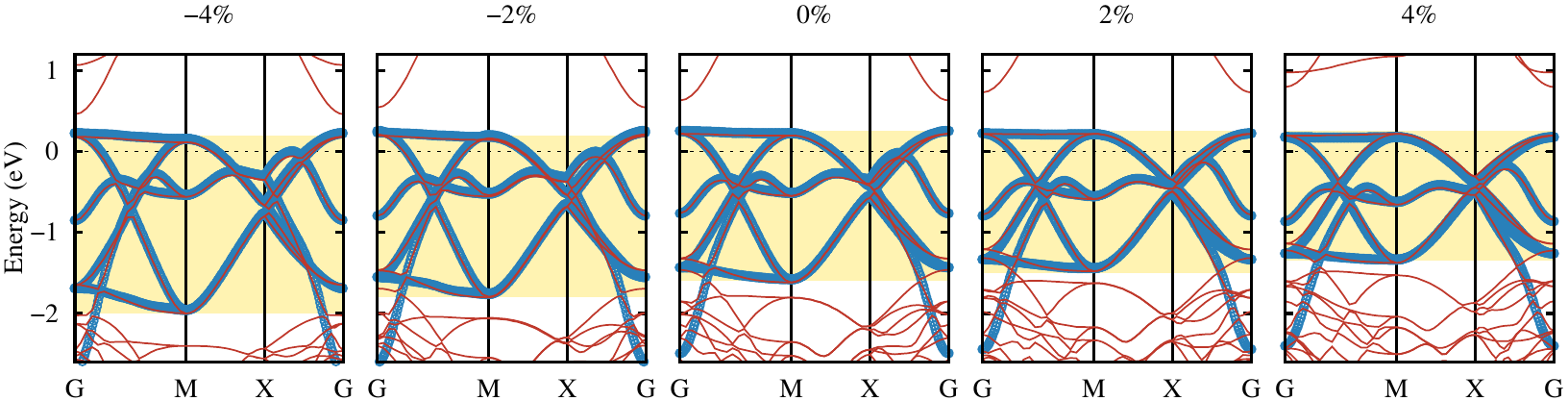}
\caption{(Color online)
Same as Fig.~\ref{fig2} but for $m=1$ for different strains.
}
\label{fig3}
\end{center}
\end{figure*}
%-----------------------------------------------------------------------

To quantify $U$ and $J$ from {\it ab initio}, we performed cRPA calculations, using an Ir $t_{2g}$ basis set constructed by means of Maximally localized Wannier functions obtained by the Wannier90 code~\cite{Marzari1997,Mostofi2008,Franchini2012} at each value of $m$.
We present the results obtained for $m=1$ and $m=2$ in Table~\ref{crpa1} and ~\ref{crpa2}. For the DFT+SOC+$U$ calculations, we have employed an effective $U_{eff} = \bar{U}-\bar{J}$, where $\bar{U}$ and $\bar{J}$ are obtained by averaging the $U_{ij}$ and $J_{ij}$ matrix elements in Table~\ref{crpa1} and ~\ref{crpa2}. In Figs.~\ref{fig2} and ~\ref{fig3}, the projected Wannier-interpolated band of $t_{2g}$ states are shown together with the PBE band structure as a function of $m$ and strain. To quantify the bandwidth $W$ we have taken the width of the $t_{2g}$ bands (denoted as shadow background in Figs.~\ref{fig2} and ~\ref{fig3}). For $m=\infty$, we used the cRPA value computed for the bulk phase SrIrO$_3$~\cite{PLiu2017}. The unscreened Coulomb parameters (${U^{bare}}$ and $J^{bare}$) for each $m$  are also listed in Table~\ref{crpa3}.

 \section{Results and discussions}

 \subsection{Phase diagram}

\begin{figure}[t]
\begin{center}
\includegraphics[width=85mm]{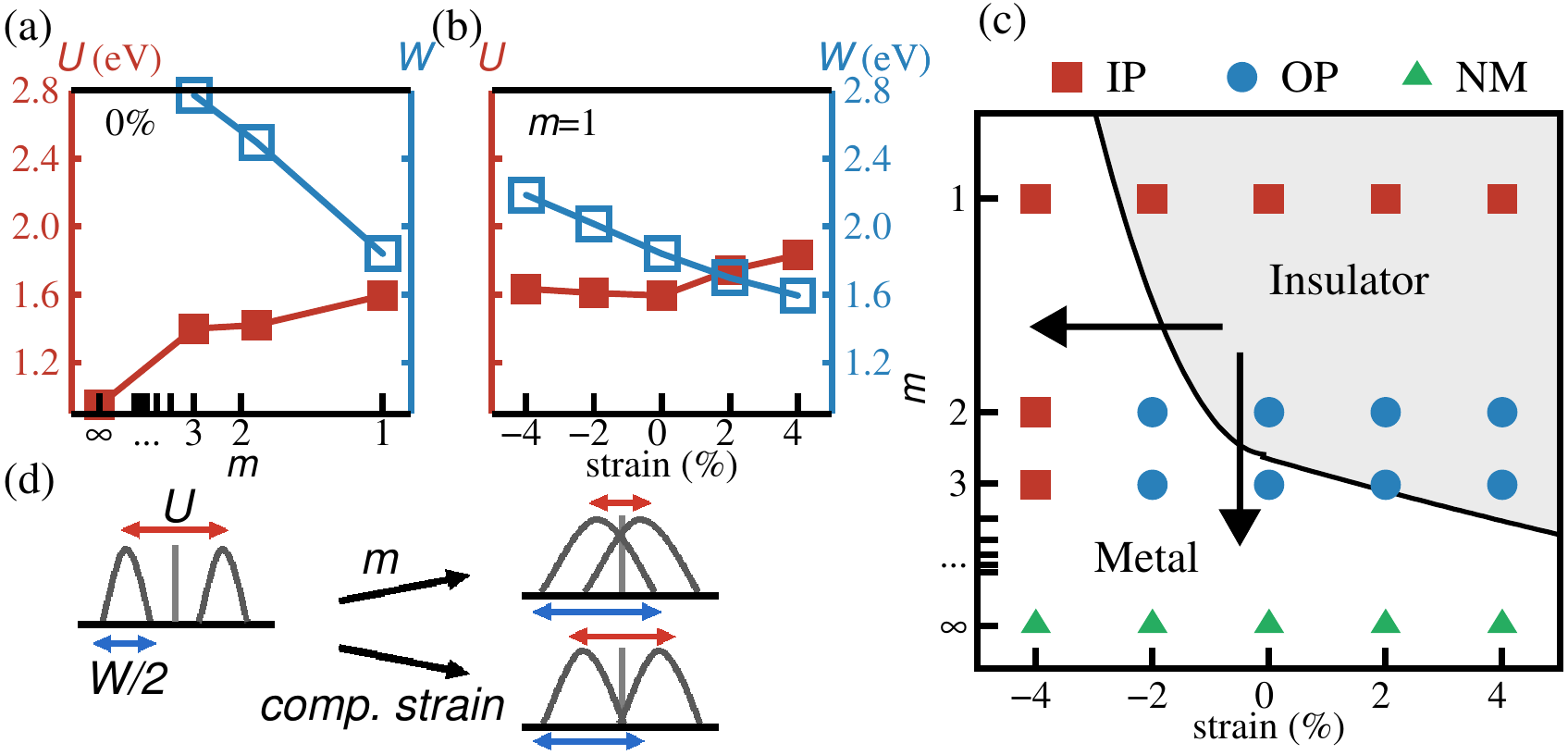}
\caption{(Color online)
Variation of $U$ and $W$ as a function of (a) dimensionality, $m$, with no strain and (b) strain (in \%) for $m=1$.
(c) Electronic and magnetic phase diagram of (SrIrO$_3$)$_m$/(SrTiO$_3$) SLs. IP and OP magnetic orderings are denoted with
filled squares (red) and filled circle (blue), respectively, whereas the nonmagnetic phase is labelled with filled triangles (green). The insulating
and metallic phases are highlighted with the gray and white background area. The arrows trace the IMT driven by dimensionality and strain. For $m=\infty$ (SrIrO$_3$) the system remains a non-magnetic metal at any strain~\cite{JLiu2013,BKim2016}. (d) Schematic band diagram showing the role of $U$ and $W$ in the IMT.
}
\label{fig4}
\end{center}
\end{figure}

 We start by inspecting the evolution of $U$ as a function of $m$ and strain (Fig.~\ref{fig4}(a)-(b)). For $m=1$ SL at the SrTiO$_3$ (STO) lattice parameter the effective $U$ for Ir is 1.59~eV, almost unchanged with respect to the corresponding value of bulk Sr$_2$IrO$_4$~\cite{PLiu2015}. As the system evolves from 2D ($m=1$) to 3D ($m=\infty$), $U$ undergoes a significant change from 1.59~eV ($m=1$) to 0.95~eV ($m=\infty$) (Fig.~\ref{fig4}(a)). Concomitantly $W$, defined here as the width of the full $t_{2g}$ in the nonmagnetic state, goes through a huge change from 1.9~eV ($m=1$) to 2.8~eV ($m=3$). Such a large reduction of $U$ ($\sim 40\%$) has been overlooked in previous studies where the dimensionality-induced ({\it i.e.} increasing $m$) IMT was interpreted solely in terms of a gradual increase of $W$~\cite{Moon2008,Matsuno2015}. However, optical conductivity measurements of the RP strontium iridates series suggest a cooperative interplay of both $W$ and $U$ across the IMT, manifested by an overall broadening of the lower and upper Hubbard bands (larger $W$) and by a shift of the characteristic $\alpha$-peak towards lower energy (smaller $U$) for progressively larger $m$~\cite{Moon2008}. This interpretation is consistent with our cRPA calculations which clearly indicate an active role of both $W$ and $U$ in the observed IMT. Since the {\it unscreened} (bare) Coulomb interaction remains almost unchanged for all $m$ in our calculations (Table~\ref{crpa3}), the dimensionality-induced reduction of $U$ is primarily due to an enhancement of the screening. This is the result of the increased coordination of each Ir site with increasing $m$ that provides further hopping channels (see Fig.\ref{fig1}(a)). For the $m=1$ quasi-2D limit $U$ and $W$ have a similar strength, comparable to the SOC energy (about 0.5~eV/Ir), and this gives rise to the wealth of intricate phenomena observed in Sr$_2$IrO$_4$~\cite{BJKim2008,BJKim2009,Nichols2013,YKKim2014,delaTorre2015,Cao2016}; however, for larger $m$, $W$ becomes the leading energy scale and pushes the systems towards a metallic and nonmagnetic regime, as represented in the phase diagram of Fig.\ref{fig4}(c).

%-----------------------------------------------------------------------
%-----------------------------------------------------------------------

\begin{figure}[t]
\begin{center}
\includegraphics[width=85mm]{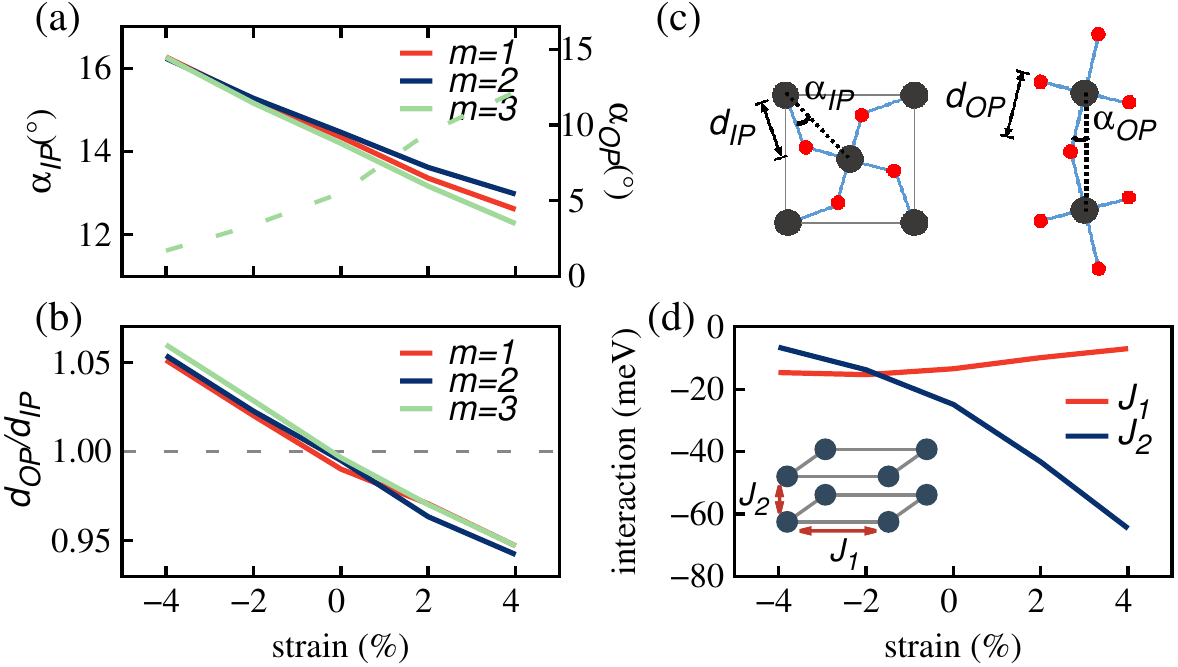}
\caption{(Color online)
(a) Change of rotation/tilting angle upon strain. Solid/Dotted lines refer to IP/OP cases. Note that $\alpha_{OP}$ is nonzero only for $m=3$ case (right axis).
(b) Change of the $d_{IP}/d_{OP}$ ratio upon strain.
(c) Schematic description of the Ir-O-Ir bond angle and Ir-O bond length both in the $ab$ plane ($\alpha_{IP}$ , $d_{IP}$) and in the out-of-plane ($\alpha_{OP}$, $d_{OP}$) direction.
(d) Change of $J_1$ and $J_2$ as a function of strain. (Inset) Schematic structure of the Ir sublattice for $m=2$, showing the IP and OP exchange interactions $J_1$ and $J_2$, respectively~\cite{u1}.
}
\label{fig5}
\end{center}
\end{figure}

 Now we turn to the role of epitaxial strain.
 Following the experimental setup of Matsuno and coworkers~\cite{Matsuno2015} we have chosen STO as the reference substrate (strain = 0\%),
 which guarantees a minimal lattice mismatch with the known members of the RP series (Fig.~\ref{fig1}(c)).
 We have considered realistic substrate-induced compressive and tensile strain up to $\pm$ 4\% with respect to the STO substrate, thus simulating the effect of different substrates, as indicated in Fig.~\ref{fig1}(c).
 The changes of $U$ and $W$ upon strain for $m=1$, reported in Fig.~\ref{fig4} (b), are more modest than those caused by varying $m$: the value of $W$ for the most compressive case (-4\%) is about 37\% larger compared to the 4\% tensile strain case, while the change in $U$ is only about 10\% and remains essentially unchanged in the strain range -4\%-0\% (see Fig.~\ref{fig4}). The reason for this \emph{moderate} change of $U$ and $W$ is that strain, unlike $m$, does not modify the number of hopping channels within the SL. Rather, the effect of strain is manifested by a change of the structural distortions within the octahedra network, specifically on the planar (IP) and apical (OP) Ir-O-Ir angles ($\alpha_{IP/OP}$) and Ir-O bondlength ($d_{IP/OP}$), as shown in Fig.~\ref{fig5} (a)-(c). Like many other perovskite-type oxides, upon tensile strain $d_{IP}$($d_{OP}$) increases (decreases), whereas the octahedral rotation angle $\alpha_{IP}$ ($\alpha_{OP}$) decreases (increases) (Fig.~\ref{fig5}(a) and (b)). We predict that the continuous application of compressive strain guides the system to a IMT, which is caused by an increment of $W$. The small increase of $U$ at the most tensile limits should be attributed to the narrowing of the $t_{2g}$ orbitals (smaller $W$), which is understandable in terms of a reduction of the hopping amplitude. Having a more itinerant character, the $m=2$ and $3$ structures are expected to be less sensitive to strain effects. The different ways in which $m$ and strain induces the IMT in strontium iridates is schematically shown in Fig.~\ref{fig4} (d): for compressive strain the leading factor is the increase of $W$ at almost constant $U$; on the other hand, the tuning of $m$ affects both $W$ and $U$ since the change of dimensionality alters not only the structure but also the electronic connectivity between the octahedra.

%-----------------------------------------------------------------------
%-----------------------------------------------------------------------

 Dimensionality and strain have a strong effect also on the magnetic properties of strontium iridates. For the single layer system ($m=1$), the magnetic exchange interactions are purely 2D as in the corresponding RP compound Sr$_2$IrO$_4$, and the IP arrangement remains the most favorable one at any strain (see Fig.~\ref{fig4} (c)). For multilayer systems ($m=2$ and $3$), however, the magnetic ordering is subjected to a spin-flop transition from OP-to-IP ordering for compressive strain larger than 3\%. This transition can be explained by the dependence of the effective intra- and interlayer interaction parameters $J_1$ and $J_2$, that we have computed assuming a classical Heisenberg-type spin Hamiltonian.
 The results are shown in Fig.~\ref{fig5} (d). %employing $m=2$ system.
 For strain larger than -3\%, $J_1$ becomes the dominant magnetic interaction, as in the case of single layer system, causing a reorientation of the spin from OP to IP. $J_2$ is much more susceptible to strain than $J_1$ owing to the associated structural changes (Fig.~\ref{fig5} (a) and (b)): for $m=2$, $\alpha_{OP}$ is always zero regardless of strain but the apical Ir-O distance $d_{OP}$ decreases monotonically with progressive expansion; conversely, $J_1$ changes very little as a result of the balance between $d_{IP}$ and $\alpha_{IP}$ that follows an opposite trend upon strain.

%-----------------------------------------------------------------------
\begin{figure}[b]
\begin{center}
\includegraphics[width=85mm]{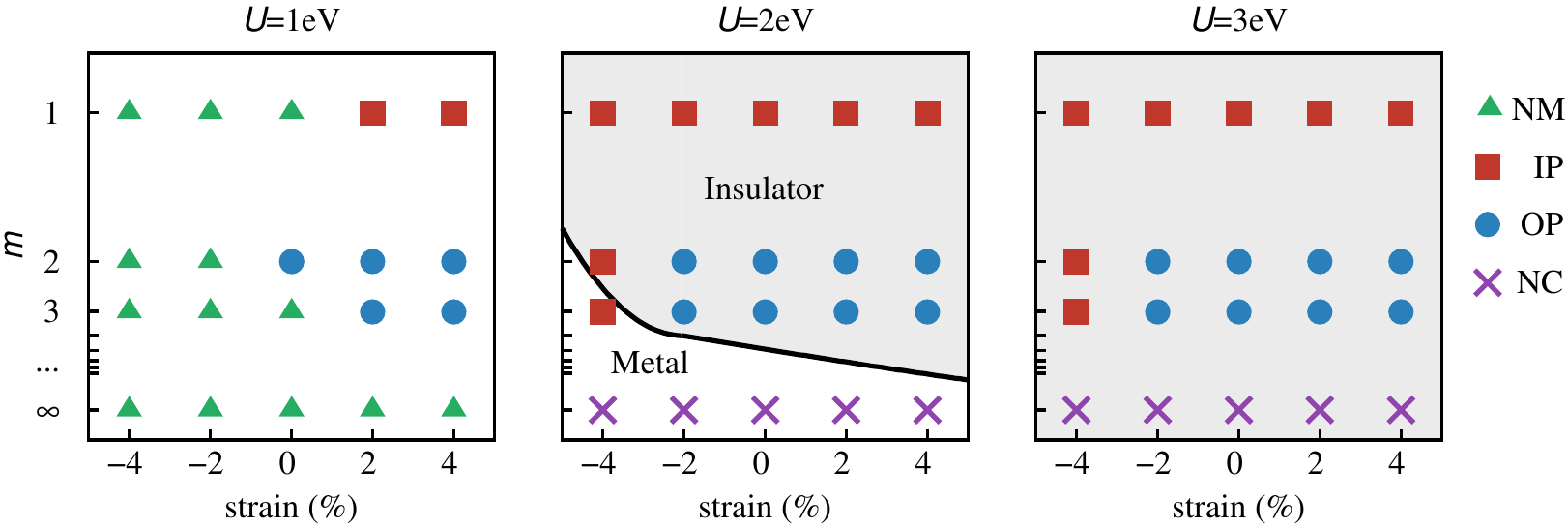}
\caption{
(Color online) Phase diagram for fixed $U$ parameters of 1~eV, 2~eV, and 3~eV. Due to the delicate balance of the $U$ with other energy scales, the overall metallicity depends highly on $U$ parameters. $U$=1~eV and $U$=3~eV give metallic and insulating phase for all $m$ values and strain ranges, respectively.
}
\label{fig6}
\end{center}
\end{figure}
%-----------------------------------------------------------------------

 By collecting all data on the evolution of the electronic and magnetic properties of strontium iridates SLs as a function of dimensionality and
 strain we have built a comprehensive dimensionality-strain phase diagram, shown in  Fig.~\ref{fig4} (c). To this purpose we have used the cRPA values of $U$ at zero strain 1.6~eV ($m=1$) and 1.4~eV ($m=2,3$) and considered IP, OP and nonmagnetic (NM) ordering. The phase diagram indicates that strain and dimensionality represent two workable means to induce two type of transitions: (i) IMT and (ii) spin-flop. The IMT can be induced either by increasing $m$ (up$\rightarrow$down) or by compressive strain (right$\rightarrow$left), and involves the two fundamentally different mechanisms sketched in Fig.~\ref{fig4} (d): cooperative $U/W$ driven IMT ($m$) and $W$-driven IMT (strain); the spin-flop transition, on the other hand, is always explained as a reduction of the intra-layer exchange interaction $J_2$ which can be tuned by a reduction of the intralyer distance $d_{OP}$ (strain) or by controlling the number of IrO$_2$ layers ($m$). We would like to underline the importance of an accurate evaluation of $U$ in the construction of the phase diagram: employing the same $U$ for different $m$, as commonly done, not only fails to describe the experimentally observed IMT from $m=2$ to $m=3$~\cite{Matsuno2015}, but also leads to the stabilization of the wrong magnetic structure. This is shown in Fig. \ref{fig6}, where we show the dimensionality-strain phase diagram obtained for $U$=1, 2, and 3~eV.
 The phase diagrams are very different to the optimum one shown in Fig.~\ref{fig4}(c).
 In particular, $U$=1~eV and $U$=3~eV give a single metallic and insulating phase, respectively, without any electronic transition.
 For the weak $U$=1~eV regime, the reduction of electronic localization produces a strong tendency towards nonmagnetic solutions.
 In the strong $U$=3~eV limit, on the contrary, the systems does not exhibit any nonmagnetic phase.
 In the intermediate limit, $U$=2~eV, the phase diagram appears overall correct, but it fails in describing some key features observed
 experimentally: (i) IMT from $m=2$ to $m=3$ is not reproduced. (ii) In the $m=\infty$, the system shows a nonvanishing magnetism
 characterized by a noncollinear (NC) magnetic structure, in disagreement with observation and with the optimum phase diagram.
 This is known to emerge when too strong correlation is employed~\cite{Zeb2012}.

 Overall, our predictions at the optimum $U$ values agree well with previous experiments on (SrIrO$_3$)$_m$/(SrTiO$_3$) SLs,
 specifically on the dimensionality-induced IMT from $m=2$ to $m=3$~\cite{Matsuno2015}. However, no sign of spin-flop transition
 was measured experimentally for SLs with $m>1$~\cite{Matsuno2015}. As the magnetic properties of iridates are highly sensitive
 to growth conditions~\cite{Sung2016}, oxygen vacancies are easily developed which could gives magnetic response different from
 the ordered pattern~\cite{BKim2017}, we ask for exact probe, such as x-ray diffraction, to resolve this issue.

 It is also worthwhile to compare the SLs with the bulk RP counterparts.
 The observed red-shift of the characteristic $\alpha$-peak for SL compared to the bulk RP system~\cite{SYKim2016} suggests a
 reduced degree of electronic correlation in the SL. This agrees well with our cRPA results:
 The $U$ undergoes a sizable reduction from the bulk Sr$_2$IrO$_4$ (1.99~eV) to $m=1$ SL (1.59~eV)~\cite{214,PLiu2017,Arita2012}.
 Note that the unscreened $U^{bare}$ for bulk Sr$_2$IrO$_4$ is 8.81~eV, similar to the corresponding $m=1$ SL value (Table~\ref{crpa3}),
 indicating  that the origin of the reduced strength of the electronic correlation should be attributed to an increased screening.
 Regarding the response upon the epitaxial strain, there has been previous optical and transport studies for the RP
 series~\cite{Nichols2013,BKim2016,RayanSerrao2013,Biswas2014}, which are are consistent with our data on the SL system.
 Finally, our previous study on  bilayer Sr$_3$Ir$_2$O$_7$, shows a spin-flop transition upon compressive strain similar to the one observed
 in the $m=2$ SL ~\cite{BKim2017}, suggestive of a similar response to the epitaxial strain.

%-----------------------------------------------------------------------
\begin{figure}[t]
\begin{center}
\includegraphics[width=85mm]{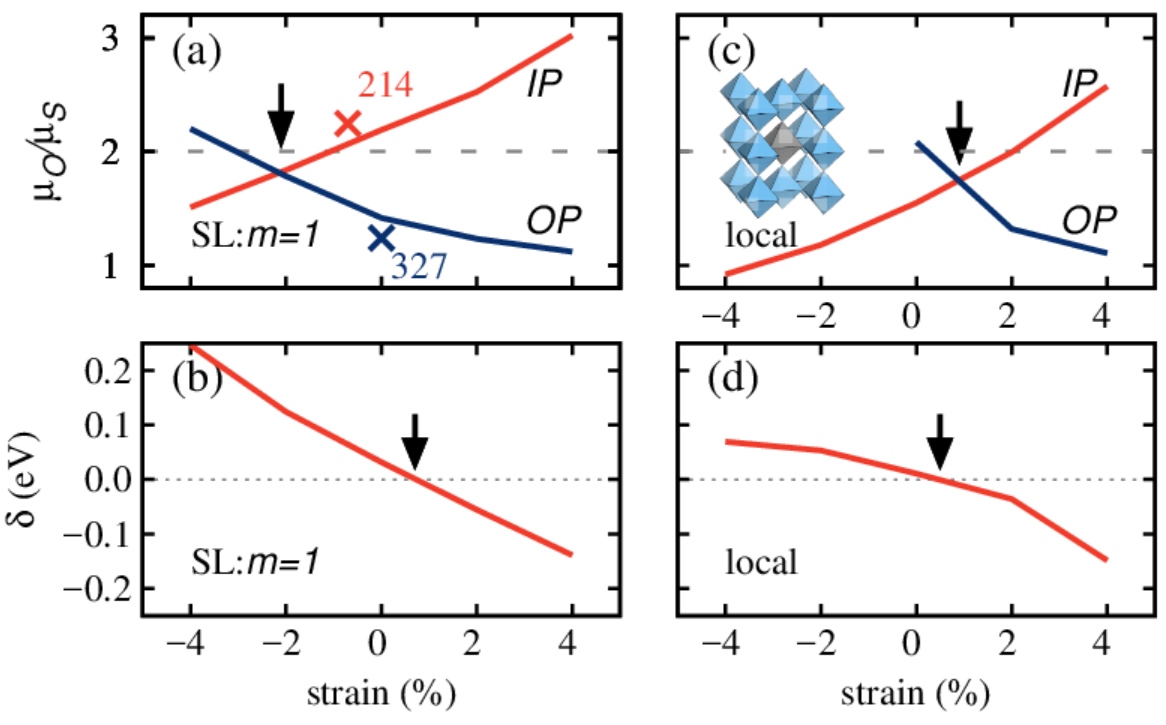}
\caption{(Color online)
(a) $\mu_O$/$\mu_S$ and (b) $\delta$ as a function of epitaxial strain for $m=1$ SL~\cite{u1}. Red and blue cross marks are for the value for bulk Sr$_2$IrO$_4$ and Sr$_3$Ir$_2$O$_7$ systems in their IP lattice parameter positions. (c) and (d) are the same for a local model.
Inset in (c) describes the structure used for the local model. The $\delta = 0$ and the $\mu_O$/$\mu_S$ =2 points are marked by arrows.
}
\label{fig7}
\end{center}
\end{figure}
%-----------------------------------------------------------------------

 \subsection{Validity of the local picture}

 Based on the above analysis, we can now study the evolution of the $J_{eff}$=1/2 and assess the validity of what is conventionally called local picture~\cite{BJKim2008,Jackeli2009}. One way to identify the $J_{eff}$=1/2-ness of a system is the ratio $\mu_{O}/\mu_{S}$ between the orbital ($\mu_{O}$) and spin ($\mu_{S}$) magnetic moment~\cite{Lado2015}. In a purely local picture the \emph{ideal} $J_{eff}$=1/2 phase associated with a cubic symmetry exhibits an isotropic $\mu_{O}/\mu_{S}$=2~\cite{BJKim2008,Jackeli2009}. However, in the presence of a tetragonal distortion, the deviation from the cubic symmetry lifts the degeneracy of the $t_{2g}$ manifold, results in a energy splitting ($\delta$=$\epsilon(d_{xy})-\epsilon(d_{xz/yz})$) and alters $\mu_{O}/\mu_{S}$.
 Indeed we find that within the SL geometry tensile strain induces a decrease of $\delta$ and a spin-dependent anisotropic
 bifurcation of the $\mu_{O}/\mu_{S}$ curve (Fig.~\ref{fig4}(a) and (b)).

\begin{figure}[h]
\begin{center}
\includegraphics[width=85mm]{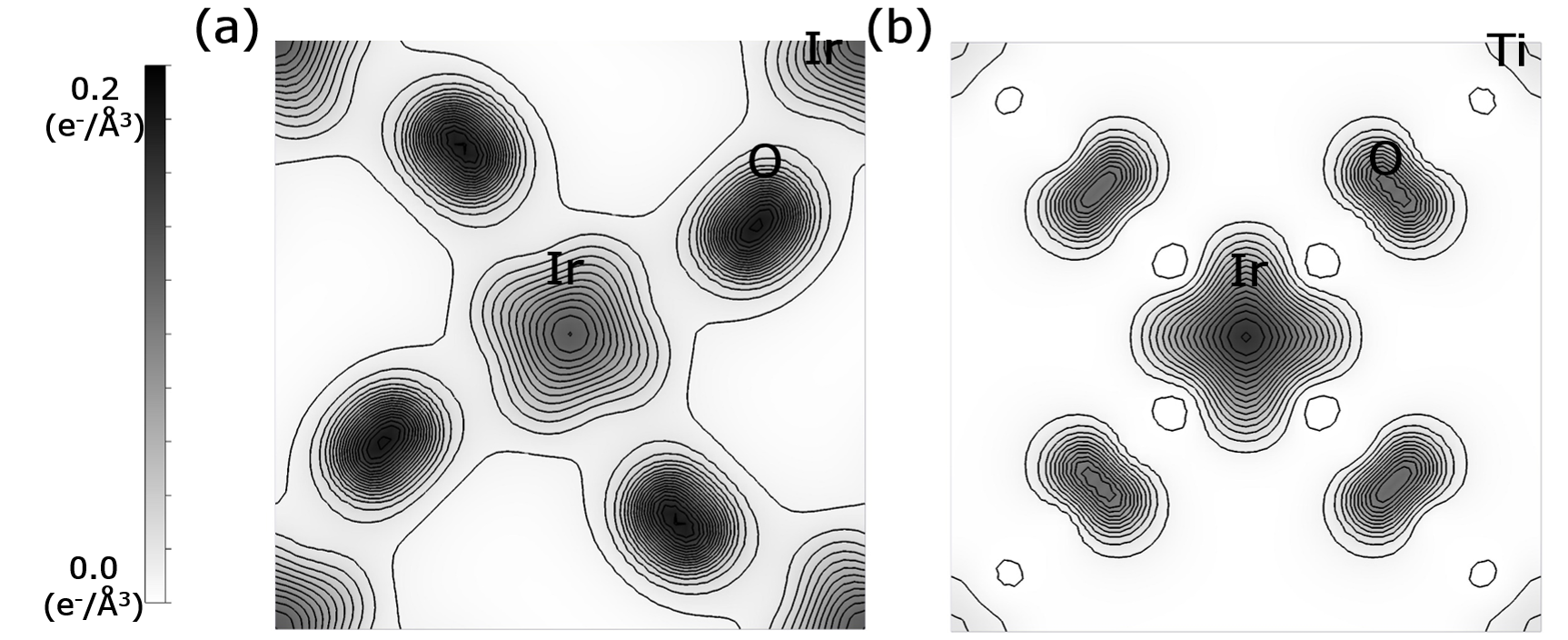}
\caption{
 Charge density plot in the IrO$_2$ plane for (a) SL $m=1$ case and (b) local structure. Both plots are obtained from the 0\% epitaxial strain.
}
\label{fig8}
\end{center}
\end{figure}

 Our findings on the evolution of $\mu_{O}/\mu_{S}$  and $\delta$ with strain conflicts with the local picture since the isotropic limit
 ($\mu_{O}/\mu_{S}$=2) found at -2\% strain (arrow in Fig.~\ref{fig7}(a)) does not coincide with the cubic limit, associated with $d_{IP}=d_{OP}$ (Fig.~\ref{fig5} (b)) and $\delta \approx 0$ (arrow in Fig.~\ref{fig7}(b)), which is instead established at about 0\% strain.
 To inspect the validity of the purely local picture we have compared the realistic SL $m=1$ situation with an ideal local structural setup constructed by isolated IrO$_6$ octahedra surrounded by TiO$_6$ octahedra as depicted in Fig.~\ref{fig7}(c). The results shown in Fig.~\ref{fig7}(c) and (d) indicate that indeed within an ideal local picture both $J_{eff}$=1/2 conditions, $\delta$=0 and $\mu_{O}/\mu_{S}\approx 2$ for both IP and OP alignments, are realized around zero strain. This is in contrast with the results obtained for the SL system, implying that the application of a local picture for realistic situations is deceptive and could lead to an incorrect analysis. Further support for this conclusion is provided by the larger degree of hybridization in the SL system compared to the artificial local model. In Fig.~\ref{fig8}, we show the charge density plot for SL $m=1$ and the local model. One can clearly see the stronger hybridization between Ir and O for the SL case compared to the local case.
 This indicates that hybridization effects are crucial to achieve a proper account of the bonding picture~\cite{Bogdanov2015}.

%-----------------------------------------------------------------------
%-----------------------------------------------------------------------

 Finally, we note that the values of $\mu_{O}/\mu_{S}$ for the bulk Sr$_2$IrO$_4$ and Sr$_3$Ir$_2$O$_7$ phases, 2.25 and 1.24~\cite{PLiu2016,BKim2017}, are almost identical to the SL values for the corresponding most favorable spin ordering (IP and OP, respectively) at the bulk lattice parameters as indicated by crossed marks in Fig.\ref{fig6}(a). This implies that not only hybridization effects but also spin arrangement have important implications on the degree of $J_{eff}$=1/2-ness of the system: the IP-to-OP spin-flopping perturbs strongly the $J_{eff}$=1/2 phase as manifested by the largely reduced $\mu_{O}/\mu_{S}$ ratio of 1.24 in  OP-ordered Sr$_3$Ir$_2$O$_7$ and SL-$m=1$ in comparison with the IP-ordered case~\cite{Zhang2013,King2013,BKim2017}.

\section{Conclusion}
In summary, we have presented a dimensionality-strain phase diagram of (SrIrO$_3$)$_m$/(SrTiO$_3$) superlattices that shows the tunability of the electronic and magnetic properties of the system as notably illustrated by a spin-flop and insulator-to-metal transition. Mismatch of the isotropic $J_{eff}$=1/2 phase with tetragonal distortion shows the incompleteness of the local model for this system. As we have shown, the accurate  quantification of the effective $U$, typically employed as adjustable parameter in DFT calculations, is of vital importance to interpret the intricate coupling between the various degree of freedom in action in weakly correlated relativistic oxides and to provide a credible prediction of the role of external perturbations.

%%%%%%%%%%%%%%%%%%%%%%%%
\bibliographystyle{apsrev4-1}
\bibliography{bibfile}

\begin{thebibliography}{52}
\bibitem [{u1()}]{u1}%
  \BibitemOpen
  \href@noop {} {\bibinfo  {journal} {$U$=~2eV and 3~eV are used to stabilize
  various spin confiurations. Fig.\ref{fig5}(d) is for $U$=3~eV, which
  essentially gives similar behavior for $U$=2~eV case. For the local model, we
  used $U$=5~eV to stabilize the magnetic phase.}\ }\BibitemShut {NoStop}%
\bibitem [{214()}]{214}%
  \BibitemOpen
  \href@noop {} {\bibinfo  {journal} {For consistency, we fully relaxed the
  structure of Sr$_2$IrO$_4$, which slightly increased the volume compared to
  the experimental one. This may affect the overestimation of $U$ of the
  system, but previous cRPA calculations consistently give effective $U$, which
  is larger than the one for SL $m=1$ case}\ }\BibitemShut {NoStop}%
\end{thebibliography}

\begin{acknowledgments}
B.K. thanks J. Matsuno and V. V. Shankar for fruitful discussions. This work was supported by the joint Austrian Science Fund (FWF) and Indian Department of Science and Technology (DST) project INDOX (I1490-N19). P.L. is grateful to the China Scholarship Council (CSC)-FWF Scholarship Program. Computing time at the Vienna Scientific Cluster is greatly acknowledged.
\end{acknowledgments}

%%%%%%%%%%%%%%%%%%%%%%%%%%%%%%%%%%%%%%%%%%%%%%%%%%%%%%%%%%%%%%%%%%
%%%%%%%%%%%%%%%%%%%%%%%%%%%%%%%%%%%%%%%%%%%%%%%%%%%%%%%%%%%%%%%%%%

%\vspace*{250px}
%\newpage

\end{document}